\documentclass[]{spie} 
\usepackage[]{graphicx}

\title{QUASI-AUTOMATIC SOFTWARE SUPPORT FOR GAIA GROUND BASED OPTICAL TRACKING} 

\author{S. Bouquillon\supit{a}, C. Barache\supit{a}, T. Carlucci\supit{a}, F. Taris\supit{a}, M. Altmann\supit{a,b},  A. H. Andrei\supit{a,c,d,e}, R. Smart\supit{c}, I. A. Steele\supit{f} and S. Els \supit{g}
\skiplinehalf
\supit{a}Observatoire de Paris (SyRTE), 77 av. Denfert-Rochereau, Paris (75014), France; \\
\supit{b}Zentrum für Astronomie der Universität Heidelberg / ARI / Germany; \\
\supit{c}Osservatorio Astrofisico di Torino / INAF / Italy; \\
\supit{d}Observatorio Nacional / MCTI / Brazil; \\
\supit{e}Observatorio de Valongo / UFRJ / Brazil; \\
\supit{f}Liverpool John Moores University / ARI / UK; \\
\supit{g}European Space Astronomy Centre / ESAC / Spain
}

\authorinfo{Further author information: (Send correspondence to S\' ebastien Bouquillon)\\S\' ebastien Bouquillon : E-mail: sebastien.bouquillon@obspm.fr}

\begin{document} 
  \maketitle 

\begin{abstract}
The ESA Gaia satellite mission will create a catalog of 1 billion stars with unprecedented astrometric precision. To achieve its aim in terms of astrometric precision, a ground based optical tracking campaign (GBOT) of the satellite itself is necessary during the five years of the mission. We present an overview of the GBOT project as a whole in another contribution\cite{Altmann14} (Altmann et al. in SPIE category ``observatory operations"). The present paper will focus more specifically on the software solutions developed by the GBOT group. 
\end{abstract}



\section{INTRODUCTION}
\label{sec:intro}  

The Gaia satellite was launched the 19th December 2013. This ESA mission will realize a catalog of 1 billion stars and observe minor objects of Solar System with unprecedented astrometric precision. To achieve its aim in terms of astrometric precision, the standard procedure for satellite tracking (i.e. by a single ranging and communications station) will not be sufficient either for correcting the relativistic aberration effects or for precisely measuring the parallaxes of solar system objects. This is the reason why a Ground Based Optical Tracking campaign (GBOT) of the Gaia satellite has been set up and will be performed during the 5 years of the mission, with daily observations using optical CCD frames taken by a small network of medium class telescopes ($\sim 2$ m) located all over the world. The requirement for the absolute accuracy on the satellite position determination is 20 mas. Because this level of accuracy can not be reached with current star reference catalogs, the images will need to be re-reduced more than 2 years after the mission start when the first data from Gaia itself will be available.

	We will present an overview of the GBOT project as a whole in another contribution\cite{Altmann14} (Altmann et al. in SPIE category ``observatory operations"). This paper focuses more specifically on the three main software products developed to conduct GBOT' s activities which are: a Web interface named \textit{Field of View Maker} (allowing GBOT observers quick access to all relevant information and data for tracking the Gaia satellite), a set of accurate astrometric reduction programs named \textit{GBOT Astrometric Reduction Pipeline} (specially adapted for tracking moving objects with the use of CCD detectors), and a dedicated database to store images and reduction parameters, intermediary data and results.
 
All these tools have been already tested with asteroids and artificial satellites orbiting around the L2 point (WMAP, Planck, Hershell and more recently Gaia satellite) and will continue to be tested and optimized intensively during the 5 years of the Gaia mission.   

The first section gives a very short overview of GBOT activities to show where each software product exists. The next three sections are devoted to the presentation of each product with greater emphasis on the GBOT Astrometric Reduction Pipeline which is the cornerstone of these products. We will conclude this paper by some general comments about the possibilities to use the GBOT specific tools for other astronomical aims.

\section{GBOT ACTIVITIES OVERVIEW} 
\label{sec:AC_OV} 

As the GBOT activities are extensively described in the SPIE paper of M. {Altmann\cite{Altmann14}}, we will restrain the following section to the optical part of the GBOT and only to the description of the place and role of the three main GBOT software products in the different steps of the GBOT process.

\begin{figure}[hb]
\begin{center}
\begin{tabular}{c}
\includegraphics[height=10cm]{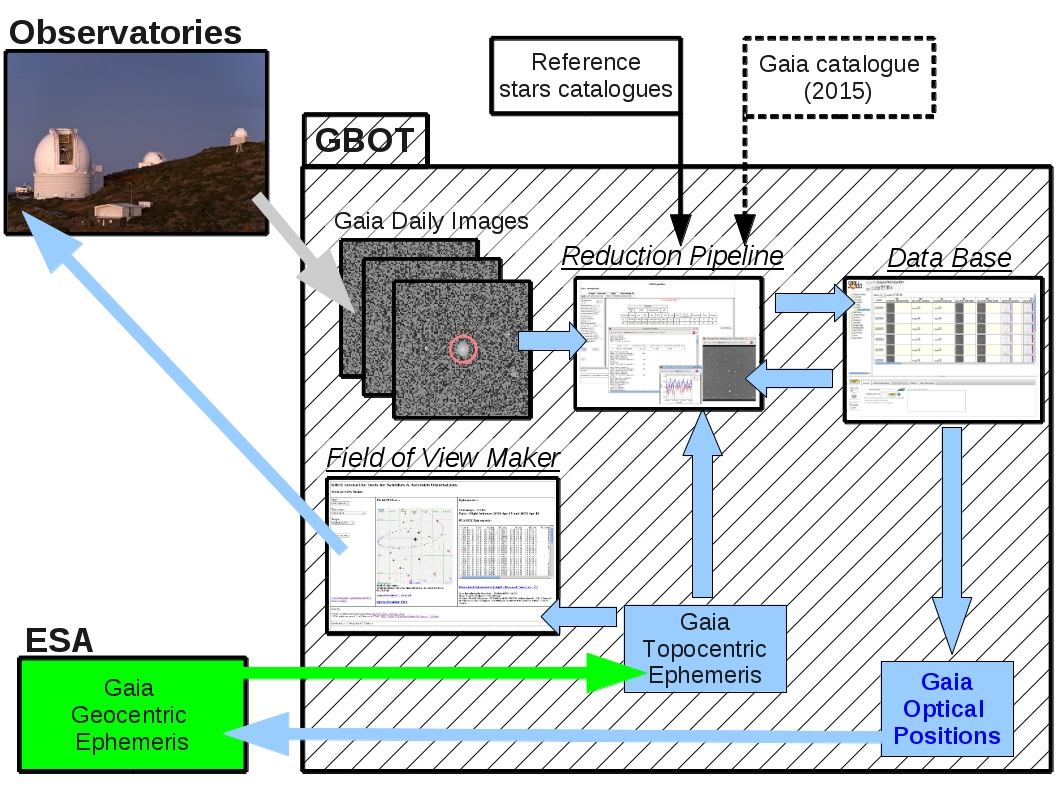}
\end{tabular}
\end{center}
\caption{ \label{fig:G_AC_OV} Overview of GBOT activities}
\end{figure}

A schematic overview of the GBOT data processing is shown in figure~\ref{fig:G_AC_OV}. It can be seen as a cyclic process.

Let us start the description of this cyclic process by the download of the target ephemeris (the current target is the Gaia satellite but could be e.g. other target in space as other artificial satellites or asteroids). For the Gaia satellite, the ephemeris is provided to us by the Gaia Mission Operation Centre (MOC) of ESOC via the Gaia Science Operation Centre (SOC) of ESAC (in ICRF and with TDB as time scale). 

The first step in the GBOT process is the creation of topocentric ephemeris (in ICRF and UTC) to provide to each collaborating observatory. This includes the prediction of the Gaia coordinates and other helpful data to facilitate the observations. A Web-browser based tool, named \textit{Field of View Maker}, allows display of this information to each observatory (see section \ref{sec:FoVM}).
 
The second step is the automatic download of the Gaia images taken by the different observatories (this process part is not detailed here) followed by a first \textit{quasi-automatic} astrometric reduction. These tasks are performed by the \textit{GBOT Astrometric Reduction Pipeline} (see section \ref{sec:GARP}).  

The third step is the storage of all the GBOT images, the results of their reduction and the related and relevant information in a dedicated and customised \textit{GBOT Database} (see section \ref{sec:GDB}).

During the five years of the Gaia mission, we will also need to frequently redo the astrometric reduction of all the images to improve and homogenize the reduction; for instance when the reduction process will be modified (enhancement of the astrometric calibration of the CCD-detectors, enhancement of the centroiding algorithms, ...) or when a new release of Gaia catalog is available and usable as reference catalog to perform new astrometric calibration of each image (the first release of Gaia catalog is expected for the end of 2015). For these reasons, and to allow a quick control of the reduction quality of each observation, we develop a Web interface allowing execution of a new astrometric reduction process for a set of observations directly selected in the database (see section \ref{sec:GDB})   

The last step in the GBOT cyclic process is the periodic delivery of all Gaia topocentric celestial coordinates deduced from the GBOT optical observations to Flight Dynamic Centre of ESOC (via SOC) to allow them to improve the Gaia ephemeris. 
 
\section{FIELD OF VIEW MAKER AND EPHEMERIS WEB INTERFACE} 
\label{sec:FoVM} 

This Web interface has been developed by Teddy Carlucci and S\' ebastien Bouquillon (SyRTE/Observatoire de Paris) to help observers to have quick access to all information and data needed to well-perform optical tracking of artificial satellites orbiting around the L2 point (as WMAP, Planck, Herschel and Gaia). In the near future, we plan to extend this tool to Asteroids.

\begin{figure}[ht]
\begin{center}
\begin{tabular}{c}
\includegraphics[height=11cm]{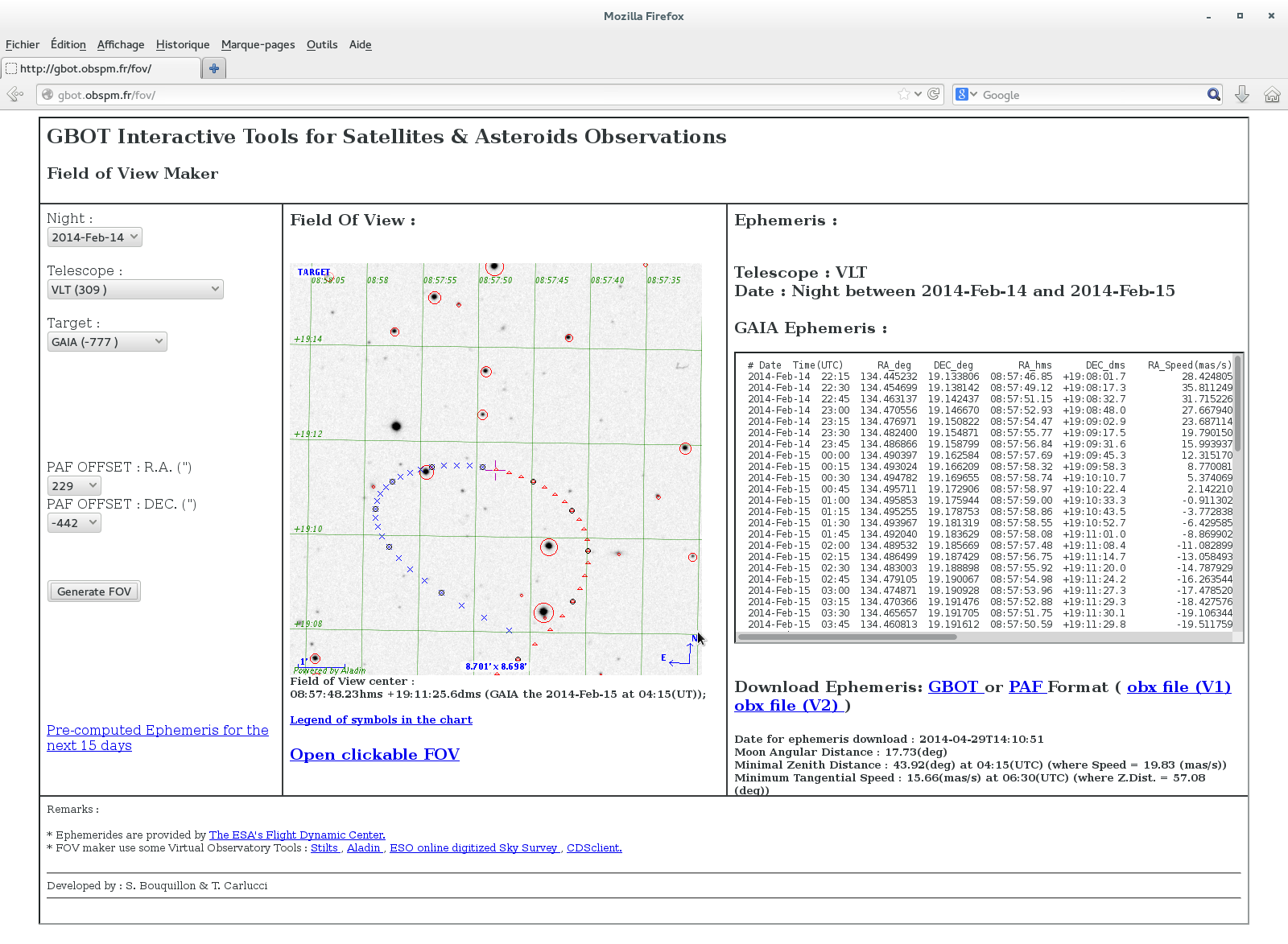}
\end{tabular}
\end{center}
\caption{ \label{fig:G_FOVM_OV} Screen snapshot of \textit{Field of View Maker} Web interface (http://gbot.obspm.fr/fov/).}
\end{figure}

  The user of this interface only needs to select the \textit{telescope}, the \textit{target} and the \textit{date}\footnote{In the future version of this tool adapted for asteroids, we plan to also add the \textit{observation time}  and the \textit{size of the field of view} as new options selectable by the user.} corresponding to his or her planned observations in the left side of the Web page (see the screen snapshot in figure~\ref{fig:G_FOVM_OV}) and to press the \textit{Generate FoV} button. 

  A field of view map (FoV Map) centered on the target position for minimum zenith distance\footnote{We choose to center the FoV Map on the target position for minimum zenith distance since this position will minimize the differential chromatic refraction and so will simplify the image reduction process.} will be displayed in the center of the screen. On this map, the main stars are indicated by red circles with radius proportional to brightness and the successive positions of the target along its trajectory with a 15 min time-frame, by blue crosses for positions before the minimal zenith distance and, after, by red triangles. A clickable map is also available with more data (the speed of the target for each point of its trajectory, zenith distance, distance to the Moon, etc.). The clickable and non-clickable FoV Maps are created when the user pushes the \textit{Generate FoV} button with the help of an automatic script using VO-tools (Aladin\cite{Aladin} and Stilts\cite{Stilts}) while the background image is downloaded from the ESO online digitized Sky Survey. 
  
  On the right side of the Web page (Figure~\ref{fig:G_FOVM_OV}) is the corresponding topocentric ephemeris of the target with all relevant data which can be downloaded in different formats by the users. 
  Some complementary information and advice are also displayed on the right bottom part of the screen concerning for instance the best moments of observation (minimal zenith distance time or minimal target speed time), or the date of the last update of the geocentric ephemeris used to compute and plot the topocentric ephemerides and the FoV Maps. 
  
  Some very specific options have also been recently implemented to facilitate the GBOT planning of observations done with VLT Survey Telescope. These options allow automatic creation of the files needed for tracking moving target with ESO telescopes: as a target ephemeris compatible with the VLT parameter file (\textit{PAF}) or the \textit{Observation Block Requirements (OB)} (Right ascension and declination offsets can even be added to the target position to shift the target in a preferred $x$ \& $y$ position in the OmegaCAM CCD-mosaic). 
   
  The routines used to make the transformation between geocentric and topocentric reference system (for ephemeris transformations) are the ones of the \textit{GBOT Astrometric Reduction Pipeline} developed in GNU fortran-95 and described in the next section.  
    
  Finally, a version of this tools for general public was developed for the launch of Gaia Satellite with some small differences: main world cities instead of professional telecopes, possibilities to choose the size of the field of view map, etc.(http://gaiainthesky.obspm.fr/fov/).  
	
\section{GBOT ASTROMETRIC REDUCTION PIPELINE} 
\label{sec:GARP}
 
  The \textit{GBOT Astrometric Reduction Pipeline} has been developed by S\' ebastien Bouquillon with the help of G\' erard Francou, Fran\c cois Taris, Christophe Barache and more generally with the inputs of all the other co-authors. This pipeline allows to derive high precision astrometry for moving celestial bodies with optical CCD detectors. Its main features are: the possibility to choose from a large set of centroiding algorithms well-adapted to moving objects, robust estimators for the error of positions, and a unique framework for all the parts of the reduction process (specifically, the control of the all of the parameters are gathered in a unique monitor file to avoid the choice of some no-coherent set of parameters between each part of the process). Another special feature of this pipeline is to not consider the images independently, but to automatically group together \textit{similar images}\footnote{The notion of \textit{similar images} is defined in details in the subsection \ref{subsec:HM}}.
  
   The pipeline is available with a sample set at http://gbot.obspm.fr/gbotpipeline/. It is based on two hundred and fifty subroutines written in GNU fortran-95 under CeCILL free software licence, deals only with FITS images and is divided into eight main routines:
\begin{itemize}
\item The \textit{HeaderModif} program which homogenizes the headers of the images and groups together \textit{similar images}.
\item The \textit{FindSources} program which detects in each image all the sources brighter than a threshold value and for each of them extracts the related data in pixel and ADU (position, scattering, flux, ...).
\item The \textit{CatalogueMaker} program which downloads the reference catalog stars corresponding to the field of view of each image.
\item The \textit{AstroReduc} program which identifies among the sources detected into each image, the stars belonging to the reference catalog and performs the astrometric (and photometric) calibration.
\item The \textit{CentImprover} program which improves the determination of the centroid of the sources (only when the centroiding method selected by the pipeline user needs a first preliminary astrometric calibration).
\item The \textit{TargetFinder} program which finds the target among the sources detected into each image and improves the determination of its centroid and other characteristics.
\item The \textit{MakeTiff} program which creates TIFF previews of the image reduction results. 
\item The \textit{MakeReductionOverview} program which gathers all the reduction parameters, results and statistical data for each set of observations (i.e. for each group of \textit{similar images}).
\end{itemize}

\begin{figure}[ht]
\begin{center}
\begin{tabular}{c}
\includegraphics[height=9cm]{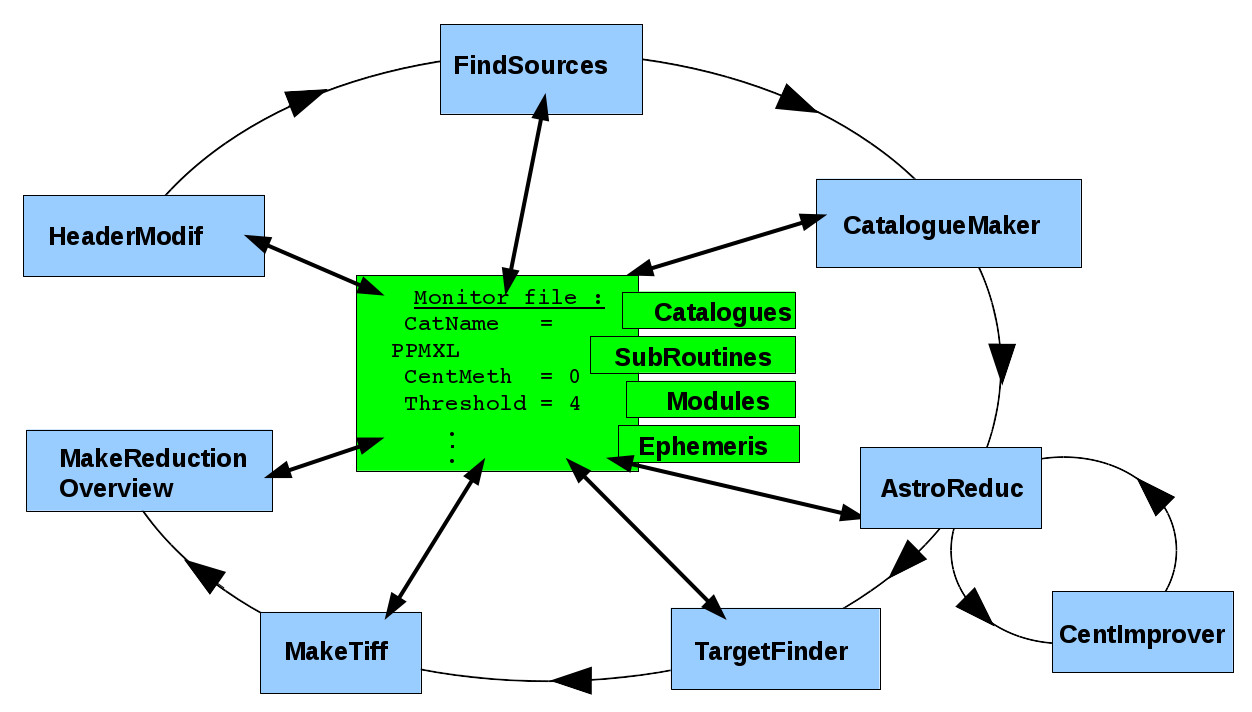}
\end{tabular}
\end{center}
\caption { \label{fig:G_RP_OV} Schematic organization of the main components of the \textit{GBOT Astrometric Reduction Pipeline} and their links}
\end{figure}

  Figure \ref{fig:G_RP_OV} gives a schematic overview of the links between these different routines and the next sections will explain with more details all these pipeline components.

\subsection{Monitor files and common subroutines}
\label{subsec:MF}
  All of these programs use the same set of reduction parameters gathered in one monitor ASCII-file to avoid the choice of  incoherent values of these parameters in the different steps of the reduction process (for instance, the PSF model selected for stars need to be coherent with the PSF model selected for the target). More details about the role of each parameter of the \textit{Monitor file} are given in the appendix \ref{sec:annA}.
  
  All of the main routines are written with the same set of subroutines for which the variables are gathered together into specific fortran modules according to the nature of each variable: one module for the variables related to reference stars,  one module for the variables related to image sources, etc. (see below). 
  
 Finally, the exchange of input and output data between the different routines of this pipeline is performed by the exchange of csv-files which can be created, completed, read or modified by each routine.
  
\subsection{\textit{HeaderModif}}
\label{subsec:HM}
  The inputs of this program are the original FITS images corresponding to different targets, taken by different telescopes and instruments at different dates. 
  
  This program has been designed to firstly gather together \textit{similar images} and sort them by date of observation. 
  The notion of \textit{similar images} is determined by the values of specific parameters selected by the user in the monitor file (see for instance in appendix \ref{sec:annA} the parameters \textit{GroupDistMax} or \textit{GroupDayMax} which allow to group together all the images with a center distances below a fixed value and all the images with observation time differences below a fixed value). 
  By default, the images considered as similar are the ones taken for the same target, with observation time delay below 1 day, with image center distance below 20 minutes of arc, taken with the same telescope, the same CCD detector and the same filter.
  
  Then the program copies each FITS image into a new FITS file with a new name according to a \textit{GBOT nomenclature}. The pixel values of the new images are identical to the original images but the header is modified by the addition of a set of \textit{GBOT-keywords} that will be extensively used by all the other programs of the pipeline. 

The header keywords of the input images are kept in the new header but as comments (except mandatory keywords which are kept as in the original header). 
Three sorts of GBOT-keywords are added:
\begin{itemize}
\item Keywords which can not be determined without the help of the keywords of the original image header as the telescope reference, the filter used or the exposure time.

\item Keywords which are mainly determined with the help of the original keywords but which can be determined by the pipeline itself as the image center in right ascension and declination or the Meteorological data from other sources.  

\item Keywords determined by the pipeline itself, such the target position and speed, the zenith distance, the Moon's angular distance or the image quality FWHM.

\end{itemize}

\subsection{\textit{FindSources}}
\label{subsec:FS}
  The inputs of this program are the FITS images modified by the \textit{HeaderModif} program described above. 
  
  This program allows detection in each image of all the sources brighter than a threshold value and for each of them to extract the related data in pixel and ADU (position, scattering, flux, ...). The different steps of this program are :
  
\begin{itemize}
\item The loading of the FITS images keywords and pixel values.

\item The construction of the pixel-masks (bad pixels, edges width, etc.).

\item The determination of background characteristics.

\item The detection of sources\footnote{In the GBOT pipeline, the definition of ``\textit{source}'' is ``\textit{a set of at least $N$ spatially connected pixels with flux in each pixel above a threshold}. The minimal number of pixels $N$ and the threshold value are fixed in the \textit{Monitor file} by the user of the pipeline (see in appendix \ref{sec:annA} the parameters \textit{ObjectMinSize} and \textit{DetectTH}).}.

\item The determination of an optimal size for the window around each source for the fitting of the PSF model (see next step).

\item The determination of source parameters by comparison with a spread function model selected by the pipeline user (the comparison is performed by a least squares algorithm written by Yong Yu(SHAO/CAS)). At this stage of the reduction process, only point spread function models (PSFs) for not moving source are available. These PSFs are: the barycentric estimation, a model based on two one-dimensional Gaussians (the two projections along x and y axes), a model based on a two-dimensional circular symmetrical Gaussian, and two models (linear and non-linear) based on one two-dimensional elliptic Gaussian  (with or without estimation of background).
\end{itemize} 

The last step of the \textit{FindSources} program is the creation of the output-files (one file per image) which contain all the characteristics of the sources extracted from each image: the positions and scattering (in pixels along the $x$ and $y$ axes in the CCD detector frame), the total fluxes and signal to noise ratios (in ADU) and all related statistical data (standard deviations and correlations) provided by the residuals between the observed and the modelled sources PSFs. We call these output-files the \textit{Sources files} and their format is CSV with a particular header compatible with the main \textit{Virtual Observatory} Tools. 

  The user can also choose or modify the method to determine the background values, the deblending option, the detection threshold, the minimal size of a source, the PSF model, some criteria to reject a source (or not),  the saturation threshold, etc. These choices are made, as for the other programs, by modifying the corresponding options in the Monitor file (see appendix \ref{sec:annA}.)

\subsection{\textit{CatalogueMaker}}
\label{subsec:CM}
 The inputs of this program are the keywords of the FITS images modified by the \textit{HeaderModif} program and in the case where the \textit{image option} (see below) has been selected, the \textit{Sources file} of each image created by the \textit{FindSources} program. 
 
 The aim of this program is to extract, from a reference catalog selected by the pipeline user, the parameters of the stars in the vicinity of the field of view of each image. This program allows to collect the astrometric characteristics of stars from one catalog and, by cross-identification, the photometry characteristics from an other catalog. The currently available reference catalogs are 2MASS\cite{2mass}, UCAC4\cite{ucac4}, PPMXL\cite{ppmxl} and GSC2.3\cite{gsc23}. To download the reference catalog, the pipeline uses the \textit{cdsclient} package developed by the Strasbourg astronomical Data Center (CDS) (more information are available on the following Web page http://cdsarc.ustrasbg.fr/doc/cdsclient.html). 
  
  Similar to the other parts of the pipeline, the configuration of the \textit{CatalogueMaker} program by the user is done through specific parameters of the \textit{Monitor file} (in particular the two parameters \textit{AstroCatName} and \textit{PhotoCatName} allow to choose the reference catalog respectively for astrometry and photometry). A specific option, named \textit{image option} (see appendix \ref{sec:annA}), allows use of the position of stars in the first image of each set of observations as astrometric reference\footnote{the right ascensions and declinations of stars of the first image are deduced from the $x$ and $y$ positions of stars available in the corresponding \textit{Sources file} and the preliminary WCS keywords of the image.}.This option is particularly useful to study the intrinsic astrometric quality of a set of observations. 
  
The output is one CSV-file per image containing all the parameters of the reference stars which are supposed to be in the image field of view as their right ascensions, declinations, proper motions, parallaxes, brightnesses and colors. We name these output-files the \textit{Stars files}. 

\subsection{\textit{AstroReduc}}
\label{subsec:AR}
  The inputs to this program are the keywords of the FITS images modified by the \textit{HeaderModif} program, the \textit{Sources files} provided by the \textit{FindSources} program and the \textit{Stars files} provided by the \textit{CatalogueMaker} program. 
  
  The aim of this program is to identify among the sources of each image, the stars of the reference catalog and to find the polynomial functions to make the astrometric and photometric links between them (in other words, to perform an astrometric and a photometric calibration of the image in the reference system of the catalog). The different steps of this program are:

\begin{itemize}

\item The loading of image keywords, source parameters and star parameters.

\item The correction of star coordinates for parallax effect.

\item The correction of star coordinates for aberration effects (due to Earth orbital and rotational motion).

\item The correction of star coordinates for refraction. 

\item The projection of star coordinates in the coordinate system of the sources. 

\item The identification of the reference stars among the image sources (by a statistical approach of the ``triangles method'')

\item The determination of polynomial functions to make the link between the sources and the star coordinates (astrometric calibration).

\item The determination of the link between the fluxes of sources and the magnitudes of stars (photometric calibration).  

\item The computation of right ascensions, declinations and magnitudes (in the reference system of the catalog) for all sources in the images.

\end{itemize} 
 
The  last step of the \textit{AstroReduc} program is to save in the \textit{Source file} corresponding to each image all the new characteristics of the sources determined with the help of the astrometric and photometric calibrations (right ascensions, declinations, magnitudes, etc.), and the related statistical data (and the residuals when the source is a source used for the calibration). 
This program also updates the WCS keywords in the FITS images
and conserves in ASCII files (that we name \textit{Calibration files}) the main data concerning the astrometric and photometric calibration parameters of each image (polynomial coefficients, number of stars used for the calibrations, etc.).

  The pipeline user selects in the \textit{Monitor file} the degree of the polynomial function, the type of polynomial function, how the saturated sources will be used to perform the calibration, if the target can be used for the link or more generally which kind of source will be used for the final calibrations (see the corresponding parametres in the monitor description in appendix \ref{sec:annA}).
  
\subsection{\textit{CentImprover}}
\label{subsec:CI}

When the pipeline user selects a more sophisticated spread function model\footnote{For instance the models corresponding to value $9$ or $10$ of the parameter \textit{CentroidMet} defined in appendix \ref{sec:annA}.} than the ones described in subsection \ref{subsec:FS}, the pipeline needs the astrometric calibration performed by the \textit{AstroReduc} program, to compute good $x$ and $y$ positions of the sources in the CCD frame. In particular, it is the case when the telescope drive system follows a moving celestial target (instead of tracking the stars as usual). In this configuration, the stars spread functions are similar to the spread function of the moving target in the usual stellar tracking mode. In this case, the best spread function model to estimate the $x$ and $y$ positions of the stars in the CCD frame is the one described in subsection \ref{subsec:TF} for moving source with a known motion. However, as the motion of the target is known through its ephemeris in right ascension and declination, we need the astrometric calibration to deduce the motion in the CCD frame.

In these cases, as the astrometric calibration is not yet performed when the \textit{FindSources} program is run, the pipeline uses a simpler PSF model to perform this first determination of the characteristics of the sources. Then the usual steps of the reduction process continue, and after the end of the  \textit{AstroReduc} program, the \textit{CentImprover} program is run to perform a second determination of all the characteristics of the sources by fitting each one with the more elaborate spread function model selected by the user. 
The old characteristics of the sources are replaced by the new ones in the \textit{Sources files} and the \textit{AstroReduc} program is run again to perform a more precise astrometric calibration by using these updated characteristics of sources.

\subsection{\textit{TargetFinder}}
\label{subsec:TF}

  The inputs to this program are the FITS images modified by the \textit{HeaderModif} program, the \textit{Sources files} provided by the \textit{FindSources} program and the \textit{Calibration files} provided by the \textit{AstroReduc} program.
  
  The aim of this program is to find and, when necessary and feasible, to improve, the coordinates of the target in the reference system of the reference catalog. To find the target, the \textit{TargetFinder} program follows a very basic method which consists of searching in each \textit{Sources files} the sources contained in a circle of radius $R$ centered on the predicted position of the target given by the target ephemeris (if several sources are contained in this circle, the closest source is considered as the target). Other options allow further constraints on the identification of the target. As the possibility to accept (or not accept) that the target would be a source previously identified as a stars of the reference catalog or the possibility to add offsets to the ephemeris prediction (in right ascension and declination) to avoid misidentification by allowing to reduce the radius $R$ of the identification circle even in the case of quite poor ephemeris prediction.      
 
   In the case of a moving target, the \textit{TargetFinder} program also allows the improvement of the target coordinates by the implementation in the GBOT pipeline of several specific spread function models for a moving source. These specific models are based on an analytical solution of the integral with respect to time of a circular Gaussian distribution with a linear motion\footnote{The two main spread function models based on this analytical solution are the one for which the motion of the target is seen as known and given by an ephemeris of the target and  the one for which the motion is seen as unknown and fitted with the other parameters of the spread function model. The choice between these models for the target spread function is done by the pipeline user through the parameter \textit{TarCentMeth} of the \textit{Monitor file} (see the appendix \ref{sec:annA}).}
   
   Figure \ref{fig:G_RP_MOG_1} shows the shape differences between the GBOT analytical solution for the spread function of a moving target (named the \textit{Moving Circular Gaussian}) and the usual elliptic Gaussian. In particular, we see, on the plots to the right of the figure, that the residuals between the spread function of a real celestial body and the usual model (top of the figure) are considerably larger that the one with the specific GBOT model for moving target (bottom of the figure).

\begin{figure}[ht]
\begin{center}
\begin{tabular}{c}
\includegraphics[height=8cm]{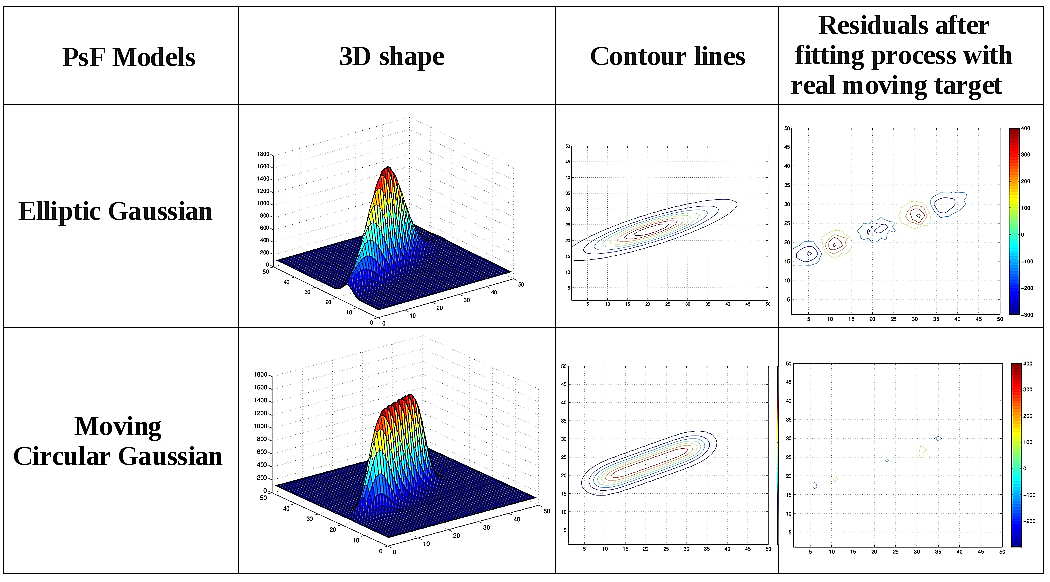}
\end{tabular}
\end{center}
\caption { \label{fig:G_RP_MOG_1} Differences between Moving Circular Gaussian and the usual elliptic Gaussian.}
\end{figure}

\begin{figure}[ht]
\begin{center}
\begin{tabular}{c}
\includegraphics[height=6cm]{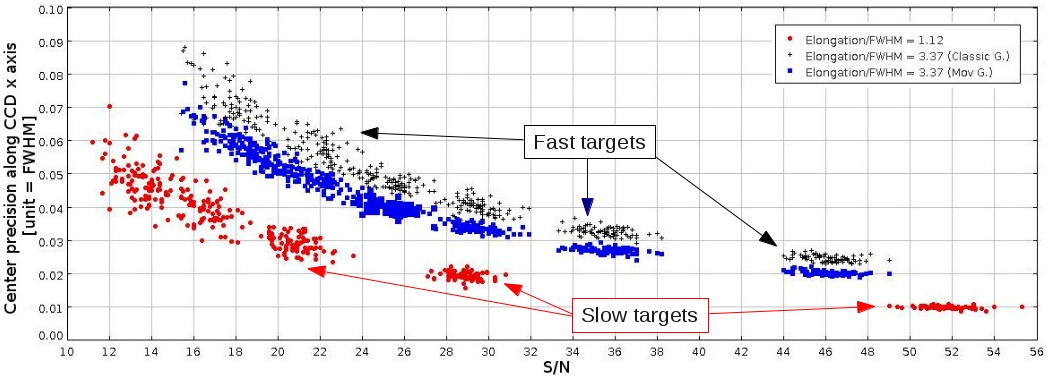}
\end{tabular}
\end{center}
\caption { \label{fig:G_RP_MOG_2} The precision of the centroiding method along $x$-axis of the CCD detector according to the source elongation (speed) and according to the source signal to noise ratio (the plot for the precision of the centroiding method along the $y$-axis is similar).}
\end{figure}

Figure \ref{fig:G_RP_MOG_2} is related to the centroiding improvement provided by the use of the specific GBOT analytical solution for the spread function of moving target according to the target signal to noise ratio ($S/N$)\footnote{We use as definition of signal to noise ratio, the definition given by Mendez in a recent publication\cite{Mendez2013}.} and to the target elongation ($L$) (which is directly related to the target speed amplitude ($V$) by the relation $L=V\times T_e$, where $T_e$ is the exposure time). This figure has been realized by running the GBOT pipeline for a set of simulated images where we can control all the parameters of the target (flux, speed, signal to noise ratio, center along $x$ and $y$ axes of the CCD detector).
The parameter plotted along the vertical axis of the figure is the standard deviation of the $x$-coordinate centre of the spread function in the CCD reference frame (obtained by the least squares method and normalized by the image FWHM).  

The dots in the lower part of the figure correspond to slow targets (corresponding to an elongation equal to $1.12$ times the image FWHM) with different values of signal to noise ratio. In this case, there is no clear improvement in the precision of the center determination provided by the use of the specific GBOT model for moving target and we plot only the results obtained by using the usual elliptic Gaussian. As expected, we see that the centroiding precision is degraded when the signal to noise ratio of the target decreases.

The other points in the upper part of the figure correspond to the precision obtained for the center determination of fast targets (corresponding to an elongation equal to $3.37$ times the image quality FWHM) with different values of signal to noise ratio and by using as spread function model respectively the GBOT specific model for moving target (squares) and the usual model (crosses). In this case, we see a clear improvement with the new model when the signal to noise ratio of the target decreases (for a same signal to noise ratio, the crosses are always above the squares and the difference increases when the  signal to noise ratio decreases). 

The last step of the \textit{TargetFinder} program is, when the target is detected among the sources, to flag the corresponding source in each \textit{Sources file} and when its coordinates can be improved, to also modify in this file all the related characteristics of this source. 

\subsection{Diagnostic data and plots}
\label{subsec:DDP}

  The aims of the last steps of the \textit{GBOT Astrometric Reduction Pipeline} are the creation of numerous diagnostic data and plots to simplify the study of the reduction quality by the pipeline user.
  
  The \textit{MakeTiff} program (based on a “fits2tiff” conversion program written by Jay Anderson (STScI)) allows, for each observation, to have visual explanations of the reduction results by creating several TIFF versions of the FITS image where, for instance, the stars used for the astrometric calibration are indicated by green circles and the target by red circle, or with a zoom centered on the target to check overlapping problem with faint stars.

  The \textit{MakeReductionOverview} program allows, for each set of observations, to gather into a unique file all the values of the parameters used for the reduction (to be able to reproduce it) and all the results concerning the astrometry and photometry of the target. The outputs are one CSV file per set of observations (named \textit{Overview file}) where each line contains the position of target, confidence parameters, magnitude, seeing, etc.
  
  To conclude, two small linux shell programs, using Stilts\cite{Stilts} and a2ps commands, allow creation of diagnostic plots (such the variation of target brightness with respect of time, or the histogram of the residuals in right ascension and declination of the stars used for the calibration) and a single PDF document which gathers all the diagnostic data and plots per set of observations.  

\section{GBOT DATABASE}
\label{sec:GDB} 

After each run of the \textit{GBOT Astrometric Reduction Pipeline}, and before the insertion of the related data in the \textit{GBOT Database}, some modifications are carried out (the transformation of CSV files into VOTable format\footnote{A VOTable format is an XML standard for the interchange of data represented as a set of tables (more details on the Web site of IVOA: http://www.ivoa.net/ ).}) as well as a basic validation process. Afterwards, the GBOT data can be inserted into the database.

This database has been built by Christophe Barache and Teddy Carlucci with the help of the SAADA freeware\cite{Michel2010} (an Astronomical Database Generator developed by L. Michel from Strasbourg observatory under GPL-license and VO-compatible). Three kind of data are inserted in the \textit{GBOT Database}: (1) the images and the data related to the observations, (2) the data and results related to the reduction processes and (3) the diagnostic data and plots. All these data are inserted in the database into three kind of formats: the FITS format, the VOTable format and the other formats (PDF, JPG, ASCII, etc.) which are gathered into the generic \textit{FLATFILE} notion of the SAADA framework. 

For the FITS format (only used for the observations), the \textit{GBOT Database} stores all the keywords of the header into the database but not the pixel values (the FITS images are kept into a repository folder and the file path stored into the database). The Web interface of the \textit{GBOT Database} displays for each FITS file all the values of the FITS keywords, the linked elements contain in the database and a JPG preview (provided in our case by the \textit{MakeTiff} program, see subsection \ref{subsec:DDP}). Several possibilities to export the original FITS images are also allowed (see below).

For the VOTable format, the \textit{GBOT Database} stores the data in a structure with two linked levels named respectively the \textit{Table level} and the \textit{Entry level}. For instance, if the VOTable format is seen as a matrix of data where each line corresponds to a set of values for several variables (columns), the lines data would be stored at the \textit{Entry level}. At the \textit{Table level} will be stored some generic data and information concerning the whole matrix: its name, its date of creation or the statistical data for each column (mean value, standard deviation, etc.). All these data (entries data and tables data) are displayed by the \textit{GBOT Database} Web interface.  
 
For the \textit{FLATFILE},  the \textit{GBOT Database} only stores the name and the path of the files, and allows through the Web interface of the database screen display of ASCII, JPG or PDF files.

The \textit{GBOT Database} can be queried by sending multicriteria requests through a user friendly control panel of the Web interface or directly by SQL scripts. The Data selected in this manner can be exported into a zip file on the user computer or, by using the \textit{WebSamp} option (the \textit{WebSampConnector} toolkit of the VO-Paris Data Centre), sent into other VO tools as Aladin\cite{Aladin} or Topcat\cite{Stilts}.

In the \textit{GBOT Database}, the data have been arranged into several collections. The main one is named \textit{SOLASTRO} and contains the \textit{Overview files} provided by the \textit{MakeReductionOverview} program (see subsection \ref{subsec:DDP}) and converted into VOTable format. 
The \textit{Table level} of this collection contains the reduction parameters used and the generic data and results concerning each set of observations (the telescope used, the filter used, the mean seeing, the mean signal to noise ratio, the reference catalog used for the calibration, the astrometric quality of each set of observations, etc.), and the links towards six other \textit{FLATFILE} collections which store plots and data at the same level (the level of the set of observations) and which are: 
\begin{itemize}
\item  The \textit{OMCPLOTS} and \textit{MAGPLOTS} collections which contain for each set of observations the plots of the variation of the target position (in right ascension and declination) and brightness respectively. 

\item  The \textit{OPTO} collection which contains the \textit{OPTO} files corresponding to each set of observations (an \textit{OPTO} file is an ASCII file which contains the essential data about the target position and magnitude).

\item  The \textit{MONITORS} collection which contains the monitor files used for each set of observations.

\item  The \textit{LOGS} collection which contains the record files of the reduction process of each set of observations.

\item  And the \textit{PDF} collection which contains the PDF documents which gathers all the data, parameters and plots produced for each set of observations.

\end{itemize}

At the \textit{Entry level} of the \textit{SOLASTRO} collection, are stored all the data and results concerning the target detection for each observation as: the date of observation, the observed $x$ and $y$ positions (and precision) of the target and the corresponding right ascension and declination, the position residual, the target magnitude, the seeing, the target signal to noise ratio, the number of stars used for the calibration, etc. All this information are displayed through the Web interface of the \textit{GBOT Database} as well as the links towards ten other collections which store data at the same level (the observation level) and which are:

\begin{itemize}
\item  The \textit{OBSERVATIONS} collection which provides access to the GBOT keywords included in the FITS image of each observation by the \textit{HeaderModif} program (see subsection \ref{subsec:HM}). 

\item  The \textit{SOURCES} collection which provides access to the \textit{Sources file} of each observation with the characteristics of all the sources extracted in the corresponding image (see subsection \ref{subsec:FS}). 

\item  The \textit{CATALOGS} collection which provides access to the \textit{Stars file} of each observation with the characteristics of all the catalogue reference stars in the vicinity of the field of view of the corresponding image (see subsection \ref{subsec:CM}). 

\item  The \textit{PREVIEWS} and \textit{TARGETZOOM} collections which provide access at a first JPG version of the FITS image where the stars used for the astrometric calibration are indicated by green circles and the target by two concentric red circles and a second JPG version with a zoom  on the target and where all the detected star are indicated by blue circles.(see subsection \ref{subsec:DDP}).

\item And the collections named  \textit{HistoOMC}, \textit{SsNxErrorBar}, \textit{ErrorBarxOMC}, \textit{RAxDECv1}  \textit{RAxDECv2}, which provided several diagnostics plots for each observations. For instance, the \textit{HistoOMC} collection provides the histogram of the residuals in right ascension and declination of the stars used for the calibration while the \textit{SsNxErrorBar} collection provides the position error bars of all the sources extracted from one image with respect to their signal to noise ratios.  
\end{itemize}
 
\subsection{A Web interface for the GBOT Reduction pipeline}

To allow a quick check of the reduction quality of each observation, Teddy Carlucci has developed a Web interface (under PHP and AJAX) for the \textit{GBOT Astrometric Reduction Pipeline} through which the reduction process can be run. This Web interface of the pipeline is connected to the Web interface of the \textit{GBOT Database} by using the \textit{delegate option} provided by the last version of SAADA.
The user accesses the Web interface of the reduction pipeline by selecting and delegating from the \textit{GBOT Database} one set of observations at the \textit{Table level} of the collection \textit{SOLASTRO}. Afterwards, several options are available for the user: display all the diagnostic plots corresponding to this set of observations, select and locally download some elements of this set of observations or run the GBOT Web Pipeline to re-reduce this set of observations.
 
\section{CONCLUSIONS}
\label{sec:conc} 
This paper has presented the software products developed in the frame of the GBOT project. A web interface, Field Of View maker, a set of accurate astrometric programs, the GBOT Astrometric Reduction Pipeline and an associated database to store images and other products during the all Gaia mission. Since the beginning of the Gaia mission, the framework is in use and has reduced and analysed the images coming from different telescopes with uncertainties around 50 mas, in good agreement with the uncertainties known for the adopted astrometric catalogs. All these software were (and are currently) developed in the frame of the Gaia mission but they are also being improved with the intention to use them for other research such as astrometric studies of other probes, detection of asteroids, or space debris.
 
\newpage
\appendix    

\section{DESCRIPTION OF MONITOR FILE PARAMETERS} \label{sec:annA}

We add in this annex the list of the reduction parameters common to all the routines and subroutines of the \textit{GBOT Astrometric Reduction Pipeline}. The values of these parameters for each specific reduction are gathered and defined in a unique ASCII file (the \textit{Monitor file}). 
\begin{table}[ht]
\begin{center}       
\begin{tabular}{|lll|}
\hline
Owner &::&The name of the person who runs the reduction pipeline.\\ 

\hline
GroupDistMax     &::& Distance maximum between center of all the images of a same set.\\
&& [ unit=seconde of arc / default value = 1200 seconds of arc ]\\

\hline
GroupDayMax	   &::& Maximal number of days between all the images date of a same set.\\
&& [ unit=day / default value =  1 day ]\\		      

\hline
PosTarImaOREph   &::& Flag for the origin of the target position and speed.\\
&&\begin{tabular}{lll}
0 (default) &:& Target position and speed give by the FITS-header\\
1 	 &:& Target position and speed give by the ephemeris
\end{tabular}\\    

\hline
TarPosType	   &::& Flag for the type of the target position in the ephemeris file.\\   
&&\begin{tabular}{lll}
0 (default)&:& Target position are topocentric \& astrometric   \\
1&:& Target position are geocentric \& astrometric\\    
2&:& Target position are geocentric \& astrometric (MOC Format 1)\\    
3&:& Target position are geocentric \& astrometric (MOC Format 2)\\	\end{tabular}\\

\hline
RefractionModel  &::& Refraction model.\\			       
&&\begin{tabular}{lll}
1 (default)&:& Laplace model\\	      
2 	 &:& Wallace model  
\end{tabular}\\								      

\hline
FLocal &::& Flag to make (or not make) the transformation between ICRF positions\\
&& and topocentric positions.\\ 
&&\begin{tabular}{lll}
0(default) &:& Do not\\
1 &:& Do
\end{tabular}\\
				
\hline
BGMethod &::& Background method.\\
&&\begin{tabular}{lll}
0(default) &:& Mean\\
1 &:& Median\\
2 &:& Mode\\
3 &:& Background map
\end{tabular}\\
&&In the case \textit{3} a baground map is used instead of a constant values.\\
&& Possible to give as input parameters:\\
&& * the sampling parameter (Nsam) corresponding to  a sampling NsamxNsam pixels\\
&& * the threshold for the pixels used to build the map \\
&& The format is (for 201x201 pixels and th=3.5) :: ``BGMethod  	 = 3,201,3.5"\\
									      
\hline
FilterForDetection &::& Flag for the image convolution with a Gaussian filter (only for sources detection)\\
&&\begin{tabular}{lll}
0(default) &:& No\\
1 &:& Yes
\end{tabular}\\
&& In the case \textit{1} the name of the filter must be given as:\\ 
&& FilterForDetection   = 1, gauss\_3.0\_7x7.conv .\\
&& The filter format is similar with the sextractor filter format.\\									      
\hline
FDeblending	&::& Flag for deblending.\\
&&\begin{tabular}{lll}
0(default) &:& No\\
1 &:& Yes
\end{tabular}\\
&& In the case 1, the parameter corresponding to the minimum flux to keep \\ 
&&a source can be added as a second parameter (as a fraction over the total flux\\
&& of the not-deblended component (0.5\% by default )).\\ 
&& The format is: FDeblending	 = 1, 0.005\\						\hline
\end{tabular}
\end{center}
\end{table} 

\newpage			      
\begin{table}[ht]
\begin{center}       
\begin{tabular}{|lll|}
\hline
MEdgeWidth	   &::& Edge width in the mask[ unit=pixel ].\\									      
\hline
DetectTh	   &::& Detection threshold [ unit = sigma of  background ].\\									      
\hline
ObjectMinSize    &::& Minimal number of pixel for a source.\\
  		      
\hline
CentroidMet	   &::& Centroiding methods.\\				      
&&\begin{tabular}{lll}
0 &:& barycenter\\
1 &:& 2 x 1D-Gaussians (along x \& y)\\
2 &:& 1 x circular 2D-Gaussian\\        
3 &:& 1 x 2D-Gaussian(linearization)\\    
4 &:& 1 x 2D-Gaussian\\		      
5 &:& 1 x 2D-Gaussian (background not fitted)\\   
7 &:& Sextractor Method\\
9 &:& 1 x 2D-Gaussian with target motion\\
10 &:& 1 x 2D-Gaussian with target motion (background not fitted)
\end{tabular}\\				      
				      
\hline
SourceSelect &::& Method for sources selection.\\
&&\begin{tabular}{lll}
0 &:& No selection\\ 			      
1 &:& OmC\_flux*NB\_pixel/flux $<$ MaxFluxErr        \\
2 &:& max(scatter\_sigma\_x \& y) $>$ MinScatterErr   \\
3 &:& \textit{1} and \textit{2} 			      \\
4 &:& Sources Selected by the user (not yet implemented option)		      
\end{tabular}\\
								      
\hline
MaxFluxErr	   &::& Maximum flux-error (for option SourceSelect=\textit{1} or \textit{3}) [ unit = percentage ].\\									      
\hline
MinScatterErr    &::& Minimum scattering-error (for option SourceSelect=\textit{2} or \textit{3}) [ unit = arcsecond ].\\									      
\hline
OutEPSF	   &::& Flag for output image of effective PSF [ 0(default) = No output / 1 = Output ].\\	
							      
\hline
OutIma	   &::& Flag for ASCII output image [ 0(default) = No outut / 1= Output ].\\
							      
\hline
SatTH 	   &::& Threshold of saturation.\\
&&\begin{tabular}{lll}
0(default) &:& Determination automatic\\
ELSE &:& Value of the threshold of saturation[ unit = ADU ]
\end{tabular}\\						      

\hline
SatPix	   &::& Ways to use the saturated pixels.\\			      
&&\begin{tabular}{lll}
0 &:& used like the other pixels\\ 		      
1 &:& sources which contain saturated pixels are not extracted\\
2 &:& sources which contain saturated pixels are extracted but saturated\\ 
&& pixel are not used for centroiding (option 2 not available at this stage)       
\end{tabular}\\								      				      

\hline
OutImaMk &::& Flag for output of mask(ASCII file).\\		
&&\begin{tabular}{lll}
0(default) &:& No outut\\
1 &:& Output
\end{tabular}\\
						      
\hline
OutBgMap &::& Flag for the Creation of a FITS-file for the background map.\\
&&\begin{tabular}{lll}
0(default) &:& No outut\\
1 &:& Output
\end{tabular}\\			      
\hline
AstroCatName &::& Name of the astrometric catalog.\\
&&\begin{tabular}{lll}
2mass (default) &:& 2MASS catalog\\
ucac2/ucac3/ucac4 &:& UCAC (version 2, 3 or 4) catalog\\
ppmxl &:& PPMXL catalog\\
gsc23 &:& GSC (version 2.3) catalog\\
image &:& The positions of stars of the first image of each set\\ 
&&of observations are used as astrometric references.
\end{tabular}\\

\hline
PhotoCatName &::& Name of the photometric catalog [ as AstroCatName or gsc23 ].\\
\hline
\end{tabular}
\end{center}
\end{table} 
			      
\newpage						      
\begin{table}[ht]
\begin{center}       
\begin{tabular}{|lll|}
\hline
ImaCenterRADEC   &::& Origin of the right ascension and declination of image center (RA \& DEC)\\
&&\begin{tabular}{lll}
0 (default) &:& RA \& DEC from FITS-header\\
1  &:& RA \& DEC from target position\\
2  &:& RA \& DEC from Monitor file
\end{tabular}\\
&& In all these cases \textit{0}, \textit{1} or \textit{2}, an offset can be added:\\
&&\begin{tabular}{lll}
&*&``BGMethod  	 = 0, 115.45, 91.55'': offset of + 115.45" and + 91.55" are\\
&&added to the image RA \& DEC center given by the FITS header.\\
&*&``BGMethod  	 = 1, 115.45, 91.55": offset of + 115.45" and + 91.55" are\\
&&added to the image RA \& DEC center given by the target position.\\
&*&``BGMethod  	 = 2, 115.45, 45.55":\\
&& the center will be the RA  = 115.45 deg. and DEC =  45.55 deg.       
\end{tabular}\\
					      
\hline
LegPoly	&::& Flag for type of polynomial function. [ 0 = Classic / 1 Legendre ]. \\						      
\hline
DegPoly	&::& Degree of the polynomial function.\\		
&&\begin{tabular}{lll}
0 (default)&:& Maximal degree allowed by the number of stars in the FoV.\\
1,..,10 &:& The degree of the polynomial function (until 10).
\end{tabular}\\
							      
\hline
ThSSN &::& S/N threshold (min. \& max. values) to keep a source as a calibrator.\\
&& [ default value = 0 (no threshold) ]\\									      
\hline
ThOMCAstro &::& Astrometric OMC $\sigma$ threshold (maximum) to consider a source as calibrator.\\
&& [ unit=seconde of arc / default value = 0 (no threshold) ]\\

\hline
ThOMCPhoto &::& Photometric OMC $\sigma$ threshold (maximum) to consider a source as a calibrator.\\
&& [ unit=seconde of arc / default value = 0 (no threshold) ]\\

\hline
ThCentErr &::& Centroiding error threshold (maximum value) to consider a source as a calibrator\\
&& [ unit=seconde of arc / default value = 0 (no threshold) ]\\
			      
\hline
MaxDistCalibTar  &::& Maximal distance from the target to consider a source as a calibrator.\\
&& [ unit=seconde of arc / default value = 0 (no maximal distance) ]\\
\hline
SatStarLK	&::& Use (or not use) the saturated stars for final calibration.\\
&&\begin{tabular}{lll}
0 (default)&:& Saturated stars are not used\\
1 &:& Saturated stars are used			     
\end{tabular}\\
									      
\hline
FTarget	&::& Flag for looking for (or not looking for) target.\\
&&\begin{tabular}{lll}
0 &:& No target to find\\
1 (default) &:& Target to find 
\end{tabular}\\
&& In the case \textit{1}, there is the possibility to add the target name.\\
&& The format is: ``FTarget	    = 1, GAIA".\\

\hline
FTarStar &::& Flag for distinguish target from catalog stars.\\      &&\begin{tabular}{lll}
0 (default) &:& The target is not a star of the reference catalog\\
1  &:& The target is a star of the reference catalog
\end{tabular}\\

\hline
TDistMax &::& Maximal distance for target. [ unit=second of arc ]	\\      	
\hline
TarCentMeth	&::& Target centroiding method [as the value of CentroidMet or 9 , 10, 11 or 12 ].\\     
&&\begin{tabular}{lll}
9 &:& 2D Gaussian with motion known\\
10 &:& 2D Gaussian with motion known and background not fitted\\
11 &:& 2D Gaussian with unknown motion\\
12 &:& 2D Gaussian with unknown motion and background not fitted
\end{tabular}\\	      

\hline								      
\end{tabular}
\end{center}
\end{table} 

\newpage						      
\begin{table}[ht]
\begin{center}       
\begin{tabular}{|lll|}
\hline
TarEphBiasRA &::& Target ephemeris offset in right ascension. [ unit=second of arc ]\\									      

\hline
TarEphBiasDEC &::& Target Ephemeris Offset in declination. [ unit=second of arc ] \\									      
		      
\hline
MaxAxisSize &::& Size maximum for TIFF images axis. [ unit=pixel ]\\									      
\hline
FlagCircle	   &::& Flag for plotting circles on the TIFF images.\\
&&\begin{tabular}{lll}
-1 &:& Circles for all detected sources \\      
0 &:& No Circle 			      \\
1 &:& Green circles for calibration stars     \\
2 &:& Red circle for target		      \\
3 &:& Green and red circles		      
\end{tabular}\\	      
									      
\hline
FTarImage &::& Flag for TIFF output of target image.\\	&&\begin{tabular}{lll}
0 &:& No\\
1 (default) &:& Yes
\end{tabular}\\
								      
\hline
TarImageSize &::& Length of the TIFF images of the target. [unit=pixel]\\    									      
\hline
TarImageFactor   &::& Scale factor for the TIFF images of the target.\\      								      
\hline
FormatESOC	   &::& Format for OPTO file.\\			      
&&\begin{tabular}{lll}
0 &:& No output\\					      
1 &:& Output with ESOC format (2010)\\
2 &:& Output with ESOC format (2012)\\			      
3 &:& Output with Minor planet center format (for asteroids)
\end{tabular}\\	      									      
  
\hline
SaveMonitorAsBest&::& Flag to back up the current monitor as the best monitor of this images set\\ 
&&in the BestMonitorRepository.\\        
&&\begin{tabular}{lll}
0 &:& No back-up (default)\\
1 &:& Back-up   	      
\end{tabular}\\
\hline

\end{tabular}
\end{center}
\end{table} 
									      		
\acknowledgments   
 
We would like to present our acknowledgements to CNES (French Space Agency) for funding the following presentation of this paper about Gaia GBOT software products to the SPIE Meeting.

\bibliography{report}   
\bibliographystyle{spiebib}   

\end{document}